\documentclass[twocolumn]{aa} 

\usepackage{graphicx,diagbox,amsmath,comment,xcolor,enumitem,ulem}

\begin{document}

\title{Interaction of eccentricity tide and convection}
%\subtitle{Subtitle}
\author{Jiachen Zheng\inst{1} \and Xing Wei\inst{1}\corrauth{xingwei@bnu.edu.cn}}
\institute{School of Physics and Astronomy, Beijing Normal University, Beijing, 100875, P.R.China}
\date{Received xxx}

\abstract{
{\it Aim.} We try to understand the interaction of eccentricity tide and convection in icy satellites. \\
{\it Method.} We use the numerical calculation of rotating convection with eccentricity tide in a spherical shell. \\
{\it Results.} We have the following results. (1) Energy scaling exponent about buoyancy force (or tidal force) decreases with increasing tidal force (or buoyancy force). (2) The combined tide-convection flow is not identical to superposition of two pure flows in respect of energy spectrum. (3) Tide controls large-scale of fluid motion but convection is important for small scale, and the mode-mode transition occurs at intermediate scales. (4) Tide suppresses heat transport efficiency. (5) Dissipation of kinetic energy scales as square of tidal force with a very weak dependence on buoyancy force, which has little impact on dissipation but suppresses the tidal influence on dissipation.
}

\keywords{tide -- convection -- icy satellites}

\maketitle
\nolinenumbers

\section{Introduction}

In icy satellites, tide and thermal convection coexist to regulate the long-term thermal evolution. Tidal dissipation provides an important internal heat source capable of supporting the long-term evolution of subsurface ocean (\citealt{nimmo2016ocean}). Thermal convection, driven by buoyancy arising from internal temperature gradient due to secular cooling, acts as the dominant mechanism of heat transport, redistributing angular momentum and shaping the interior thermal structure. Viscous processes including tidal heating, lateral flow, and convective overturn likely influence the deep ice shell of Europa \citep{nimmo2009geodynamics}. More generally, ocean tidal heating may help maintain primordial ocean temperature, and the tidal dissipation rate itself depends nonlinearly on the internal thermal structure \citep{chen2014tidal}, implying that tidal force and convective heat transport form a dynamically coupled system rather than independent processes.

The study of rotating convection has a long history, dated back to \citet{chandrasekhar1961}, from the onset (stationary at large Prandtl number v.s. oscillatory at small Prandtl number) to fully developed turbulence. In the context of icy satellites, interactions between convection and latitudinal density variations beneath spatially variable shells \citep{kang2023modulation}, as well as the role of tides and libration in driving vertical mixing and modifying buoyancy fields \citep{jansen2023energetic}, highlight that tidal processes can fundamentally modify the buoyancy field that sustains convection itself. Conversely, convection can also alter the efficiency of tidal dissipation through nonlinear interactions between turbulent flows and large-scale tidal motions \citep{kang2022different}. These studies collectively demonstrate that tidal and buoyancy-driven mechanisms are tightly coupled in shaping interior circulation and long-term thermal evolution.

The study of tide can be dated back to Laplace and Poincar\'e \citep{ogilvie2014tidal, lebars2015flows}. The tidal response can be decomposed into a large-scale quasi-hydrostatic deformation (equilibrium tide) and a wave-like dynamical tide. On equilibrium tide, how time-periodic tide suppresses turbulent viscosity of convection is a long-term debate between \citet{zahn1977tidal} and \citet{goldreich1977turbulent}. Zahn proposed that mixing length is calculated with tidal period such that turbulent viscosity $\nu_t$ is suppressed with a factor $\tau_{\rm tide}/\tau_{\rm conv}$, instead, Goldreich proposed that only eddies faster than tide contribute to turbulent viscosity such that $\nu_t$ is suppressed with a factor $(\tau_{\rm tide}/\tau_{\rm conv})^2$. In addition, Goodman used the perturbation method to find the suppression factor $(\tau_{\rm tide}/\tau_{\rm conv})^{5/3}$ \citep{goodman1997fast}. Numerical simulations have been performed \citep{penev2009dissipation, Vidal2020MNRAS, Vidal2020apjl, barker2021interaction, barker2022} and the recent results support Goldreich's idea. Recently \citet{terquem2026turbulent} shows that fast tide can be efficiently damped through correlations between velocity fluctuation and mean flow. Our work in this paper aims to understand this interaction of tide and convection.

In addition to tide, mechanically driven flows such as precession or libration share similarities with tidal flows, as they all involve large-scale shear and elliptical streamlines which can induce instability through the triad resonance of inertial waves \citep{kerswell1993, Vidal2023JFM}. The interaction of precession and stratification or convection in spherical geometry has been studied \citep{wei2013stratified,lin2026}, and more supercritical regime is thoroughly investigated in Cartesian geometry \citep{de2025tidal}. 

However, how eccentricity tide interacts with rotating convection in spherical geometry has not been investigated. This interaction reflects the mode-mode transition and the tidal effect on turbulent viscosity. In this paper we will try to measure this interaction by numerical calculations, though our parameter regime is far away from the realistic icy satellites. In \S\ref{sec:method} the equations and numerical method are introduced, in \S\ref{sec:result} the results are shown, and in \S\ref{sec:discussion} our results for the realistic situation are briefly discussed.

\section{Method}\label{sec:method}

We consider an incompressible Boussinesq fluid confined in a rotating spherical shell with inner radius $r_i$ and outer radius $r_o$. The system rotates at angular velocity $\boldsymbol{\Omega}$. The temperature is decomposed into conductive background state $T_b$ satisfying $\nabla^2 T_b=0$ in the absence of heat source and perturbation $\Theta$. We solve $\nabla^2 T_b=0$ to obtain $T_b(r)=(\Delta T/d)r_ir_o/r+(T_or_o-T_ir_i)/d$ where  $\Delta T=T_i-T_o$ is temperature difference between inner and outer boundaries and $d=r_o-r_i$ is shell thickness. The Boussinesq approximation addresses that density perturbation is considered only in buoyancy and $\delta\rho/\rho=-\alpha\Theta$. The gravitational acceleration is $\boldsymbol g=-Gm(r)/r^2\hat{\boldsymbol r}$ where the enclosed mass $m(r)=\rho(4/3)\pi r^3$ is proportional to radius, so that we write $\boldsymbol{g} = - g_0 (r/r_o) \hat{\boldsymbol r}$ where $g_0$ is gravitational acceleration at $r_o$. Then the dimensionless governing equations read
\begin{equation}
\frac{\partial \boldsymbol{u}}{\partial t} + \boldsymbol{u}\cdot\boldsymbol{\nabla}\boldsymbol{u} = -\nabla p + Ek\nabla^2\boldsymbol{u} + 2\boldsymbol{u}\times \hat{\boldsymbol z} + \widetilde{Ra}\,\Theta\,r\,\hat{\boldsymbol{r}},
\end{equation}
\begin{equation}
\frac{\partial\Theta}{\partial t} + \boldsymbol{u}\cdot\boldsymbol{\nabla}\Theta = \frac{r_i}{d}\frac{1}{r^2}u_r + \frac{Ek}{Pr}\nabla^2\Theta
\end{equation}
where length is normalised with $r_o$, time with $\Omega^{-1}$, velocity with $r_o\Omega$, pressure with $\rho(r_o\Omega)^2$ and temperature perturbation with $\Delta T$. The three dimensionless parameters are
\begin{equation}
Ek = \frac{\nu}{\Omega r_o^2}, \qquad \widetilde{Ra} = \frac{\alpha g_0 \Delta T}{\Omega^2 r_o}, \qquad Pr = \frac{\nu}{\kappa}
\end{equation}
where $Ek$ is Ekman number ($\nu$ viscosity and $\Omega$ rotation rate), $\widetilde{Ra}$ the rotationally modified Rayleigh number ($\alpha$ thermal expansivity and $g_0$ gravity acceleration at $r_o$) measuring buoyancy force, and $Pr$ Prandtl number ($\kappa$ thermal diffusivity). We choose the no-slip velocity and fixed temperature at inner and outer boundaries, i.e., $\boldsymbol{u}=0$ and $\Theta=0$ at $r=r_i$ and $r_o$.

We study eccentricity tide applying to synchronous icy satellites. We use an outer boundary flow to drive tide \citep{wei2018}, which is equivalent to an elliptical equilibrium tide capable of driving dynamical tide of inertial wave. For eccentricity tide, the boundary radial velocity at $r=r_o$ is given as \citep{lithwick2025enceladus}
\begin{equation}
u_r(r_o,\theta,\phi,t) = 3\xi\,Y_2^2(\theta,\phi)\cos(\omega t) + 4\xi\,Y_2^2(\theta,\phi)\sin(\omega t)
\end{equation}
where $\xi$ measures the tidal force and $\omega$ the tidal frequency. We fix the aspect ratio $r_i=0.5$. Usually the fluid dynamics is controlled by Ekman pumping at $Ek\sim10^{-3}$ but by inertial waves at $Ek\sim10^{-4}$, so that we fix  Ekman number $Ek=2\times 10^{-4}$ which is sufficiently small to induce inertial waves. Typical value of Prandtl number in icy satellites is $Pr\sim10$ \citep{hartmann2024} and we fix $Pr=1$ because computation becomes more difficult for $Pr$ farther away from 1. We fix tidal frequency $\omega=1$ for a synchronous state. Then we explore $\xi$ and $\widetilde{Ra}$ to study the interaction of tide and convection. $\widetilde{Ra}$ is chosen by the onset of convective instability and $\xi$ is chosen by the kinetic energy compared with that of purely convective flow (see the next section).

The equations are solved with the standard pseudo-spectral code \citep{hollerbach2000spectral}. Velocity of incompressible flow is decomposed with toroidal-poloidal scalars, variables are expanded with spherical harmonics on spherical surface and with Chebyshev polynomials in radial direction, time integral is carried out with predictor-corrector method, diffusion term is semi-implicit, and the nonlinear terms are calculated in physical space and then transformed to spectral space. The resolution as high as 60 Chebyshev polynomials, 60 harmonic degrees and 50 harmonic orders is used, and the convergence test is performed. The convective part of this code has been validated by many authors, and the tidal implementation with the boundary flow was validated in \citep{wei2018}.

\section{Results}\label{sec:result}

\begin{figure*}
 \centering
 \includegraphics[width=0.1\textwidth]{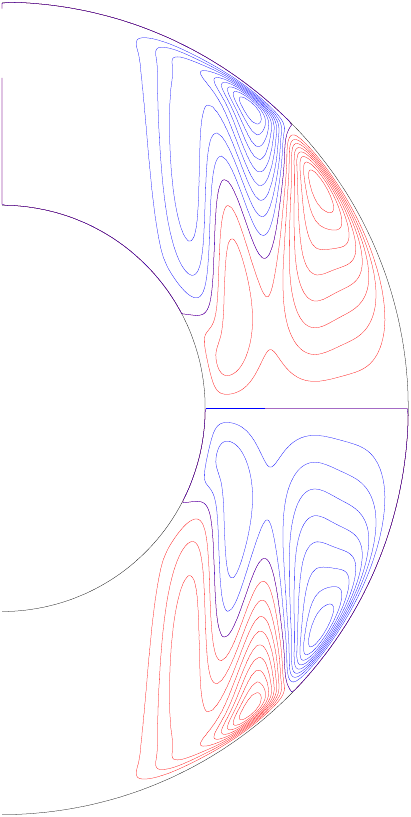}
 \includegraphics[width=0.2\textwidth]{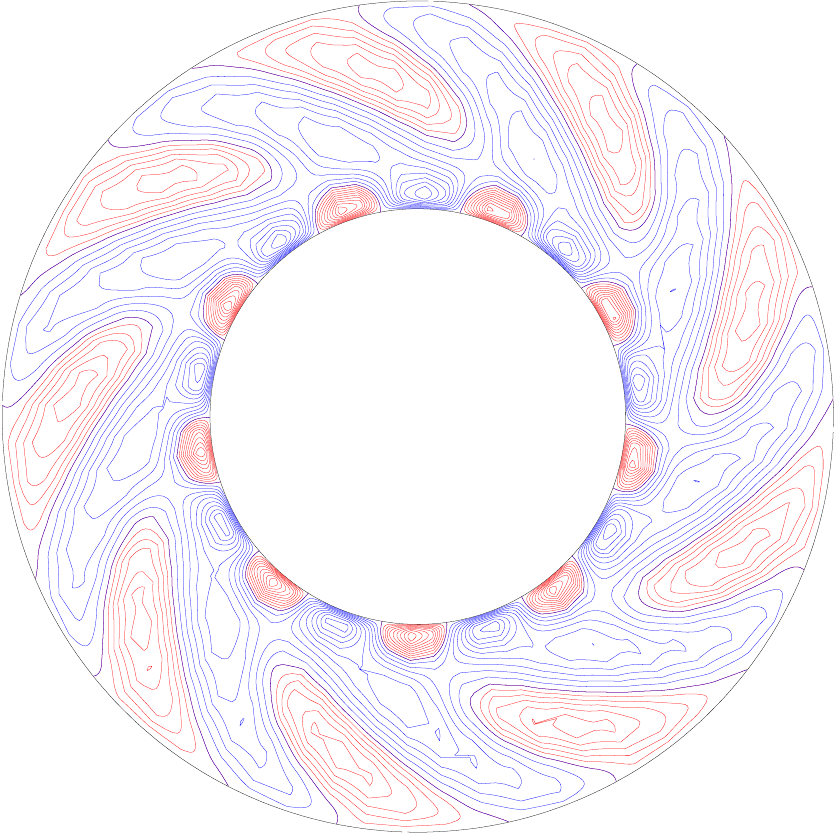}
 \includegraphics[width=0.2\textwidth]{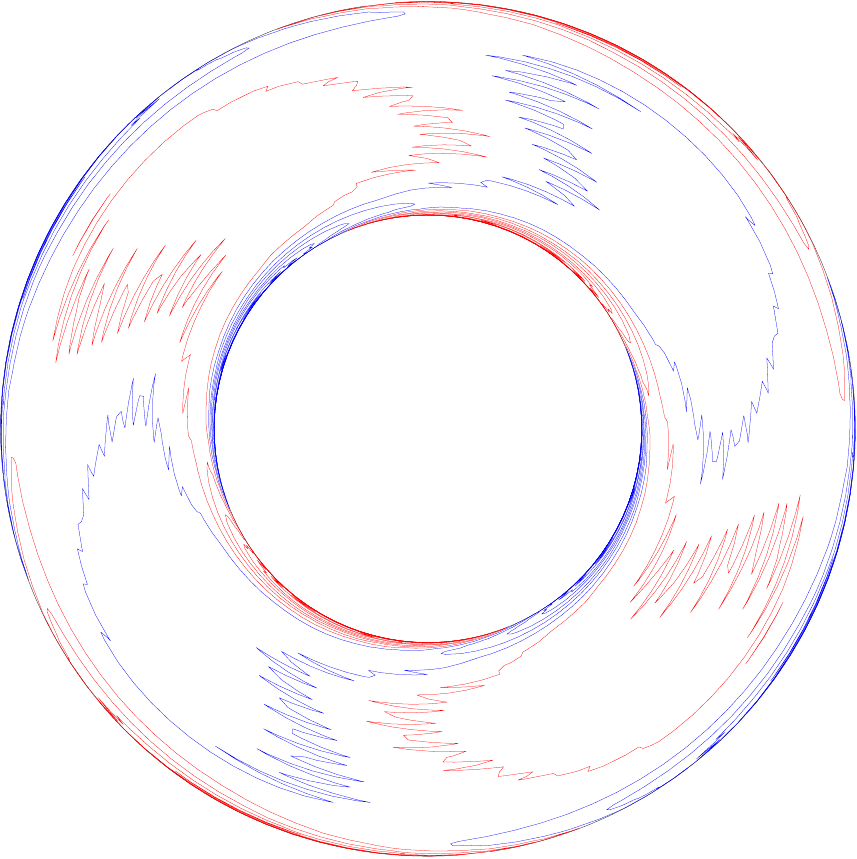}
 \caption{Left and middle panels show the meridional circulation and angular velocity in equatorial plane of convective flow at $\widetilde{Ra}=0.03$, and right panel shows the angular velocity in equatorial plane of tidal flow at $\xi = 3.90\times10^{-7}$ (the $m=0$ meridional circulation of tidal flow is tiny since tidal flow is mainly on $m=2$).}
 \label{fig:contour}
\end{figure*}

We begin by examining the flow structure in the convection-only and tide-only cases, as shown in Fig.~\ref{fig:contour}. The onset of convective instability is around at $\widetilde{Ra}=0.03$. The tidal flow at $\xi=3.90\times10^{-7}$ has almost the same time-averaged kinetic energy as that of purely convective flow at $\widetilde{Ra}=0.03$. The kinetic energy is the integral over the whole fluid volume $\int|\boldsymbol u|^2dV$. Square root of kinetic energy is the typical velocity used to define Reynolds number, which is positively correlated with buoyancy force but negatively correlated with rotation, e.g., \citep{lin2024}. We take the time-average of kinetic energy when flow reaches a statistically steady state. Convective flow exhibits an organized structure, e.g., columnar structure under fast rotation and $m=9$ spiral rolls, while tidal flow exhibits a large-scale $m=2$ structure and thin boundary layers. 

\begin{figure*}
 \centering
 \includegraphics[width=0.35\textwidth]{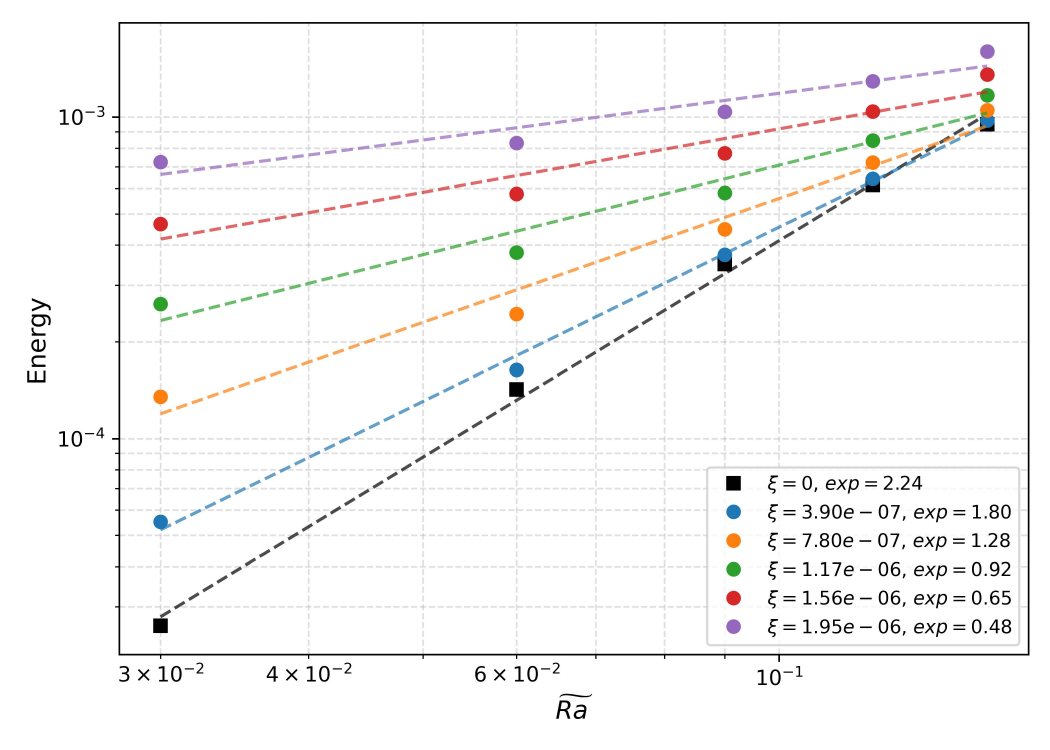}
 \includegraphics[width=0.35\textwidth]{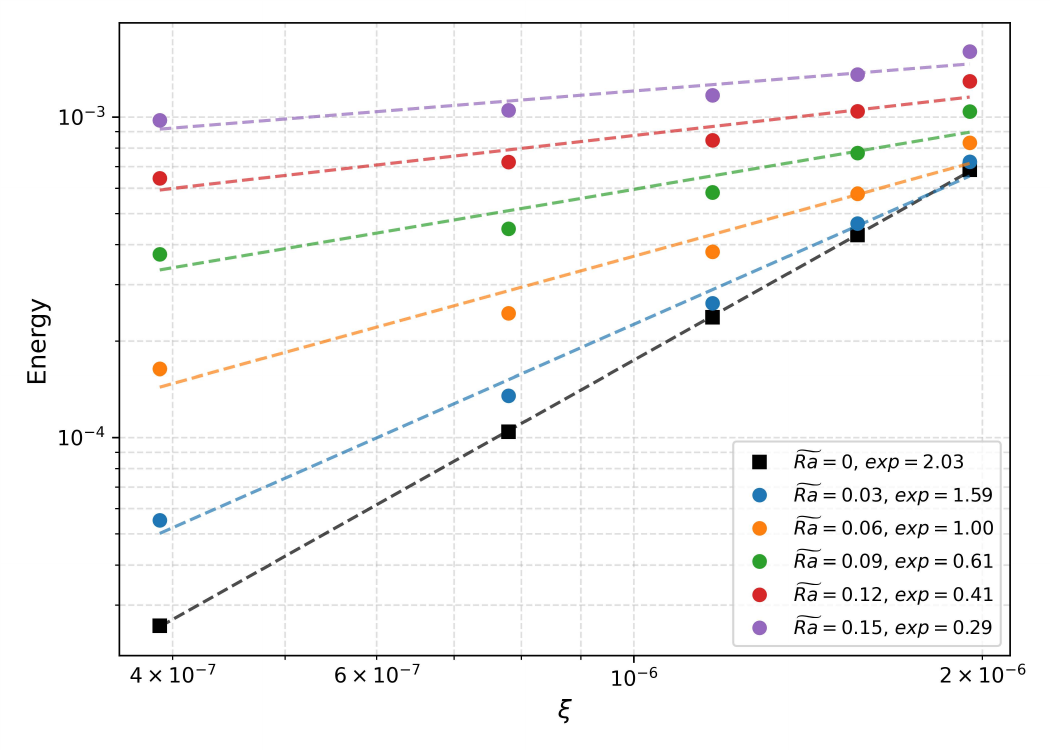}
 \caption{Time-averaged kinetic energy as a function of $\widetilde{Ra}$ at fixed $\xi$ (left panel) and as a function of $\xi$ at fixed $\widetilde{Ra}$ (right panel). In figure legends exp denote scaling exponents. The cases with $\xi=0$ and $\widetilde{Ra}=0$ correspond to convection-only and tide-only simulations, respectively.}
 \label{fig:energy_scaling}
\end{figure*}

Next we try to find the scaling law of energy. Fig.~\ref{fig:energy_scaling} shows the time-averaged kinetic energy versus $\widetilde{Ra}$ (left panel) and $\xi$ (right panel). Left panel shows that increasing tide makes the scaling exponent about buoyancy force decrease. This implies that in the presence of strong tide increasing buoyancy force may not improve fluid motion. Right panel shows the scaling exponent 2 of purely tidal flow and it decreases with increasing buoyancy force. This implies that in the presence of strong buoyancy force increasing tide may not improve fluid motion.

\begin{figure*}
 \centering
 \includegraphics[width=0.35\textwidth]{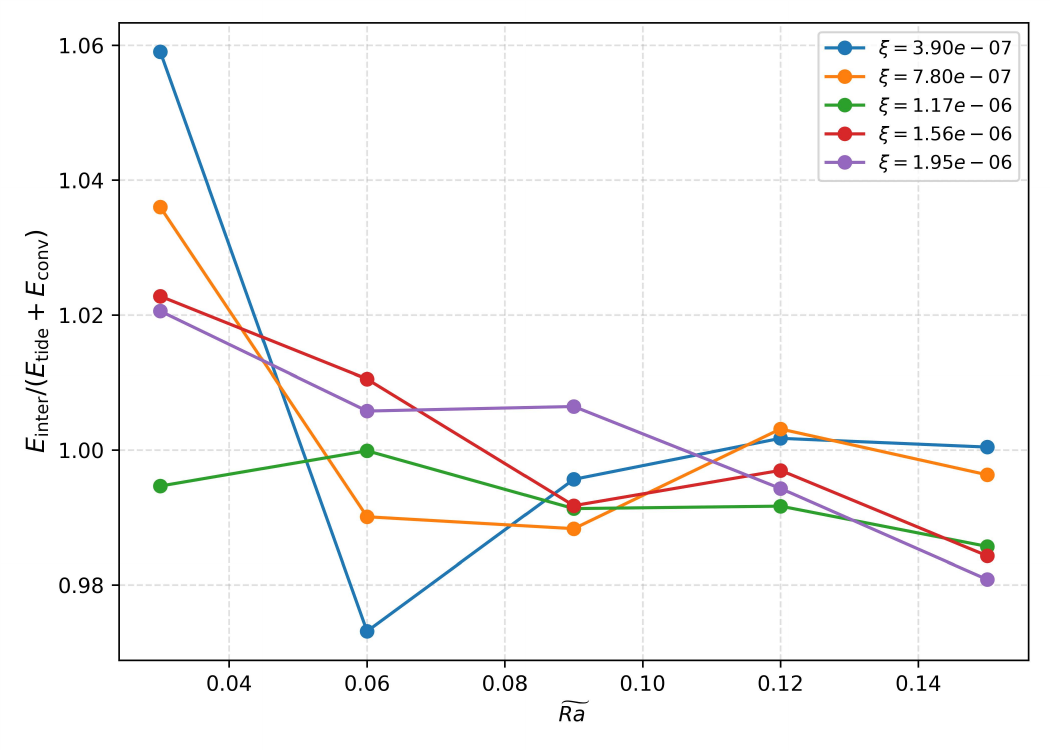}
 \includegraphics[width=0.35\textwidth]{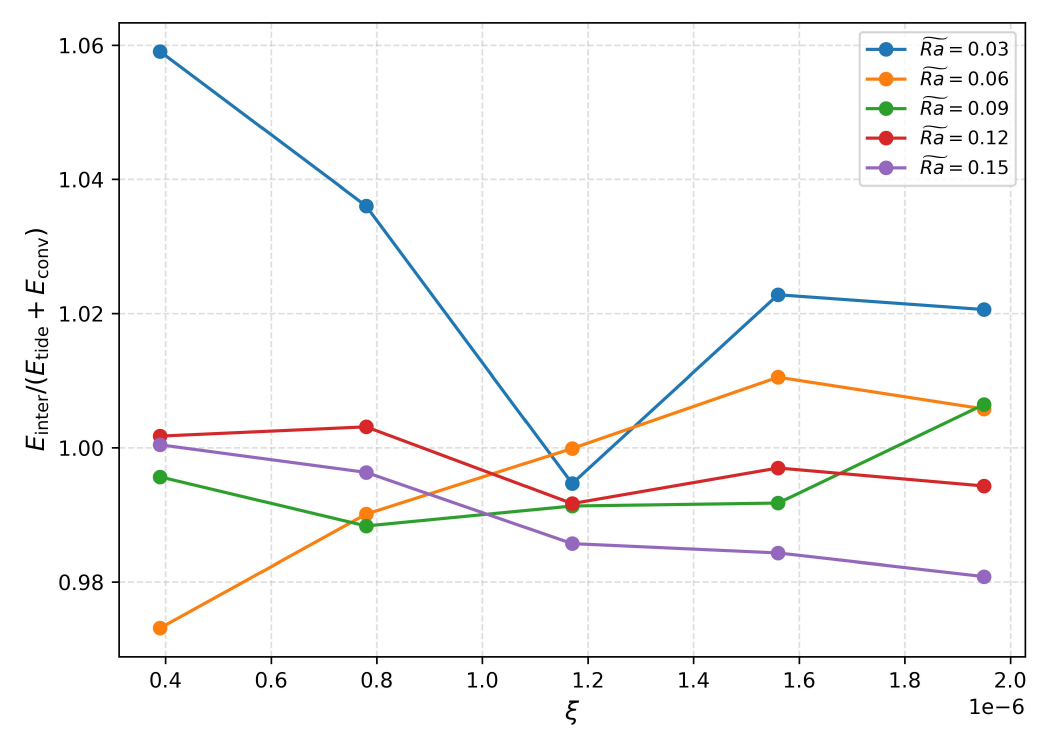}
 \caption{Interaction coefficient about energy $E_{\rm inter}/(E_{\rm tide}+E_{\rm conv})$ as a function of $\widetilde{Ra}$ (left panel) and $\xi$ (right panel).}
 \label{fig:interaction_coefficient}
\end{figure*}

Now we enter the main topic of this paper, namely the interaction of tide and convection. Fig.~\ref{fig:interaction_coefficient} shows the interaction coefficient about energy $E_{\rm inter}/(E_{\rm tide}+E_{\rm conv})$, namely the ratio of the kinetic energy of tide-convection flow to that of the sum of two pure flows. Left panel shows the coefficient versus buoyancy force and right panel versus tide. Firstly, this coefficient departs from 1, i.e., tide-convection flow is not simply the superposition of two pure flows, which indicates the interaction of tide and convection. Secondly, roughly speaking, weaker (or stronger) buoyancy force leads to a larger (or smaller) coefficient (left panel), but how the coefficient changes with tide seems chaotic (right panel).

\begin{figure*}
 \centering
 \includegraphics[width=0.75\textwidth]{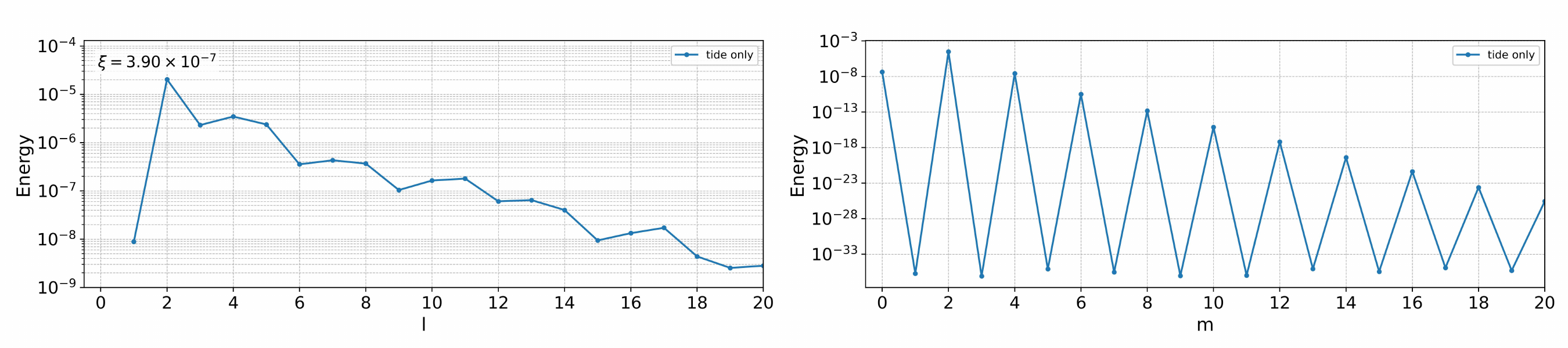}
 \includegraphics[width=0.75\textwidth]{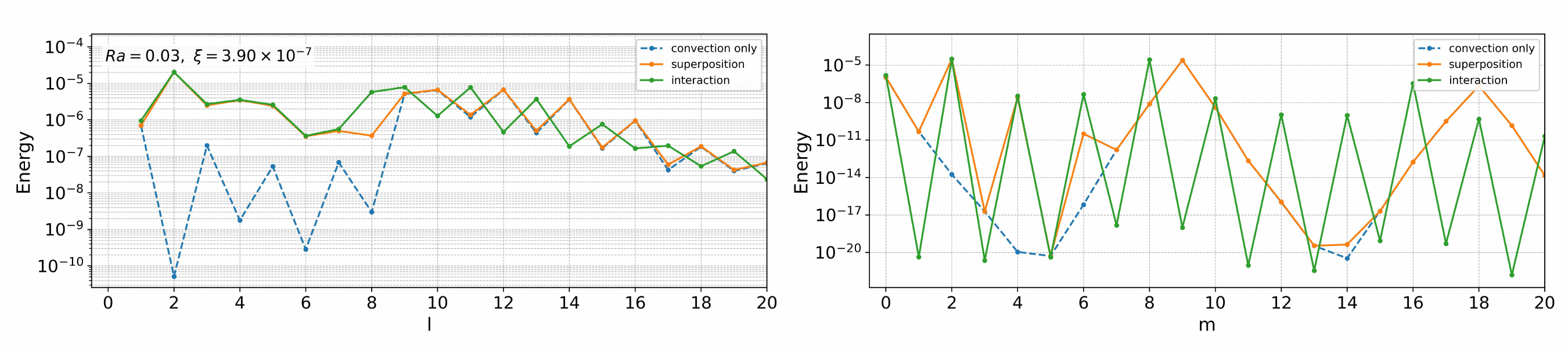}
 \includegraphics[width=0.75\textwidth]{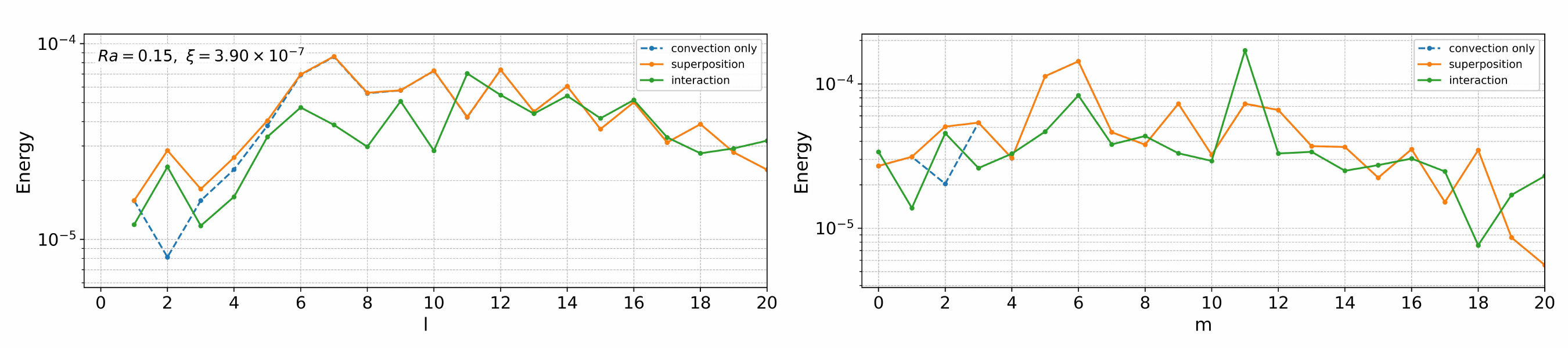}
 \includegraphics[width=0.75\textwidth]{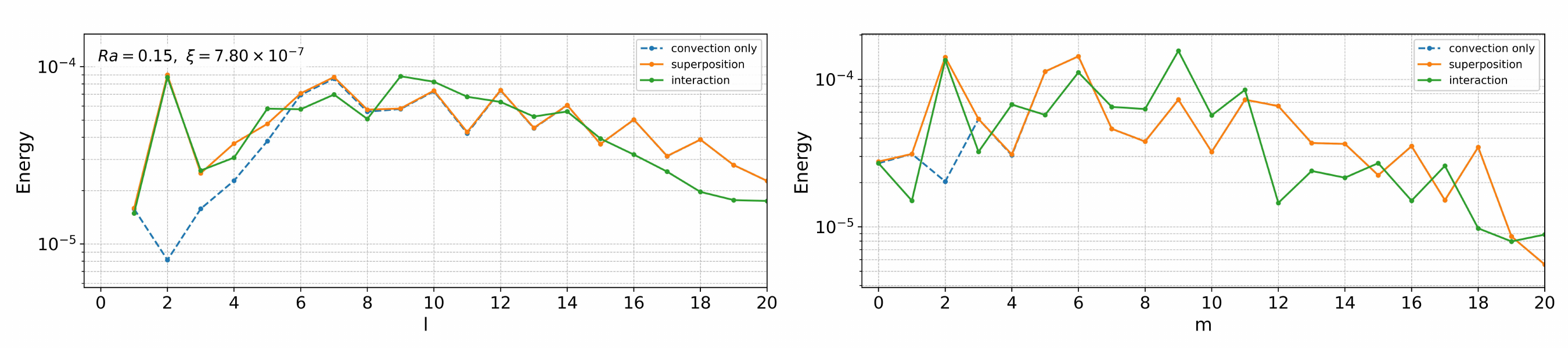}
 \caption{Energy spectra about $l$ (left column) and $m$ (right column). First to third rows: fixed $\xi = 3.90 \times 10^{-7}$ and $\widetilde{Ra}$ = 0, 0.03 and 0.15. Last row: $\xi = 7.80 \times 10^{-7}$ and $\widetilde{Ra}$ = 0.15. Purely tidal flow is denoted by solid blue curve, purely convective flow by dashed blue, the superposition of two pure flows by orange and tide-convection driven flow by green.}
 \label{fig:LMmode}
\end{figure*}

In addition to energy we investigate the energy spectra for further understanding the interaction of tide and convection. Fig.~\ref{fig:LMmode} shows the energy spectra about $l$ (left column) and $m$ (right column) of purely tidal flow, purely convective flow, the superposition of two pure flows, and tide-convection driven flow. First of all, the most striking feature is that tide-convection driven flow (green) is not equal to superposition (orange), which again indicates the interaction of tide and convection. With weak tide $\xi = 3.90 \times 10^{-7}$ and weak buoyancy force $\widetilde{Ra}=0.03$ (the second row), the two pure flows have almost equal energy. The $l$ spectrum of superposition (orange) follows tidal flow on large scale and convective flow on small scale, but that of tide-convection flow (green) follows tidal flow on large scale and neither pure flows on small scale. The $m$ spectra of superposition and tide-convection flow are similar to the $l$ spectra, and moreover, the parity of tide-convection flow follows tidal flow. With weak tide $\xi = 3.90 \times 10^{-7}$ and stronger buoyancy force $\widetilde{Ra}=0.15$ (the third row), both the $l$ and $m$ spectra show that tide-convection flow is closer to convective. The peak mode shifts to $(l=11,m=11)$, and the $m$ parity of tidal flow is broken. With stronger tide $\xi=7.80 \times 10^{-7}$ and stronger buoyancy force $\widetilde{Ra}=0.15$ (last row), tide overwhelms convection on large scale. This spectral feature implies that tide controls large scale and convection is important for small scale.

The dominant modes $(l,m)$ are shown in Table~\ref{tab:peakmode}. The convective flow usually prefers high $l$ and $m$ ($\widetilde{Ra} = 0.06$ is an exception), but the tidal flow dominates on low wavenumbers $(l=2,m=2)$. As we know, convective flow at $\widetilde{Ra} = 0.03$ and tidal flow at $\xi=3.90\times10^{-7}$ have almost equal energy, so that the fact that the tide-convection flow at $(\widetilde{Ra} = 0.03, \xi=3.90\times10^{-7})$ dominates on low mode $(l=2,m=2)$ suggests that tide is more powerful to determine flow structure than convection. In the presence of tide, kinetic energy tends to shift to large scale, which implies that mode-mode transition is likely from convection to tide. 

\begin{table*}
\centering
\caption{Dominant mode $(l,m)$ as a function of $\widetilde{Ra}$ and $\xi$. The table illustrates the transition from a convection-dominated regime at $\xi=0$ to a tide-dominated regime at larger $\xi$ where the dominant mode is $(2,2)$.
}
\label{tab:peakmode}
\begin{tabular}{c|cccccc}
\hline
\diagbox{$\xi$}{$\widetilde{Ra}$} 
& 0 & 0.03 & 0.06 & 0.09 & 0.12 & 0.15 \\
\hline
0                & $0$ & (12, 9) & (1, 0) & (12, 12) & (7, 6) & (7, 6) \\
$3.90\times10^{-7}$  & (2, 2) & (2, 2) & (2, 2) & (2, 2) & (9, 8) & (11, 11) \\
$7.80\times10^{-7}$  & (2, 2) & (2, 2) & (2, 2) & (2, 2) & (2, 2) & (2, 2) \\
$1.17\times10^{-6}$ & (2, 2) & (2, 2) & (2, 2) & (2, 2) & (2, 2) & (2, 2) \\
\hline
\end{tabular}
\end{table*}

\begin{figure*}
\centering
\includegraphics[width=0.35\textwidth]{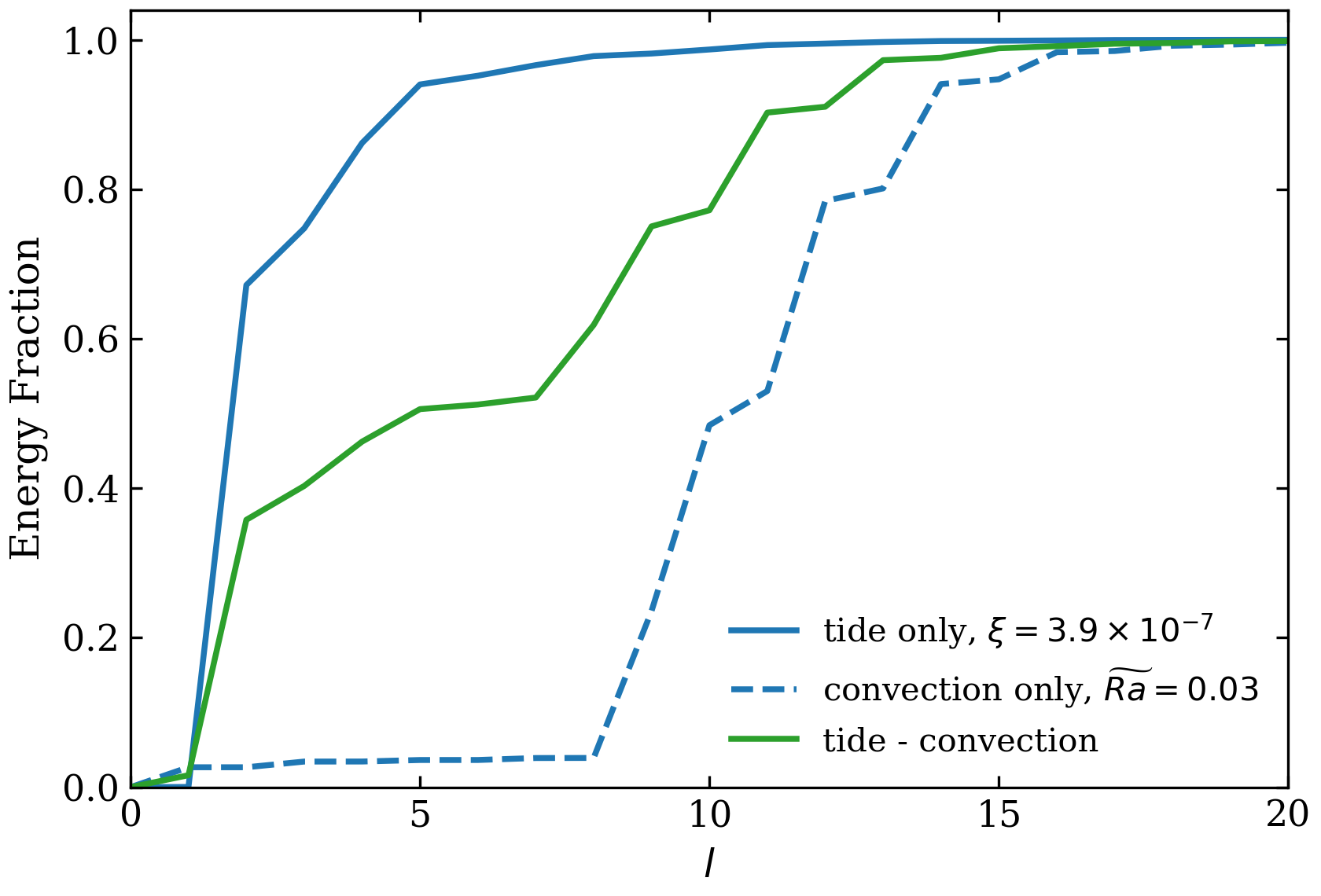}
\includegraphics[width=0.35\textwidth]{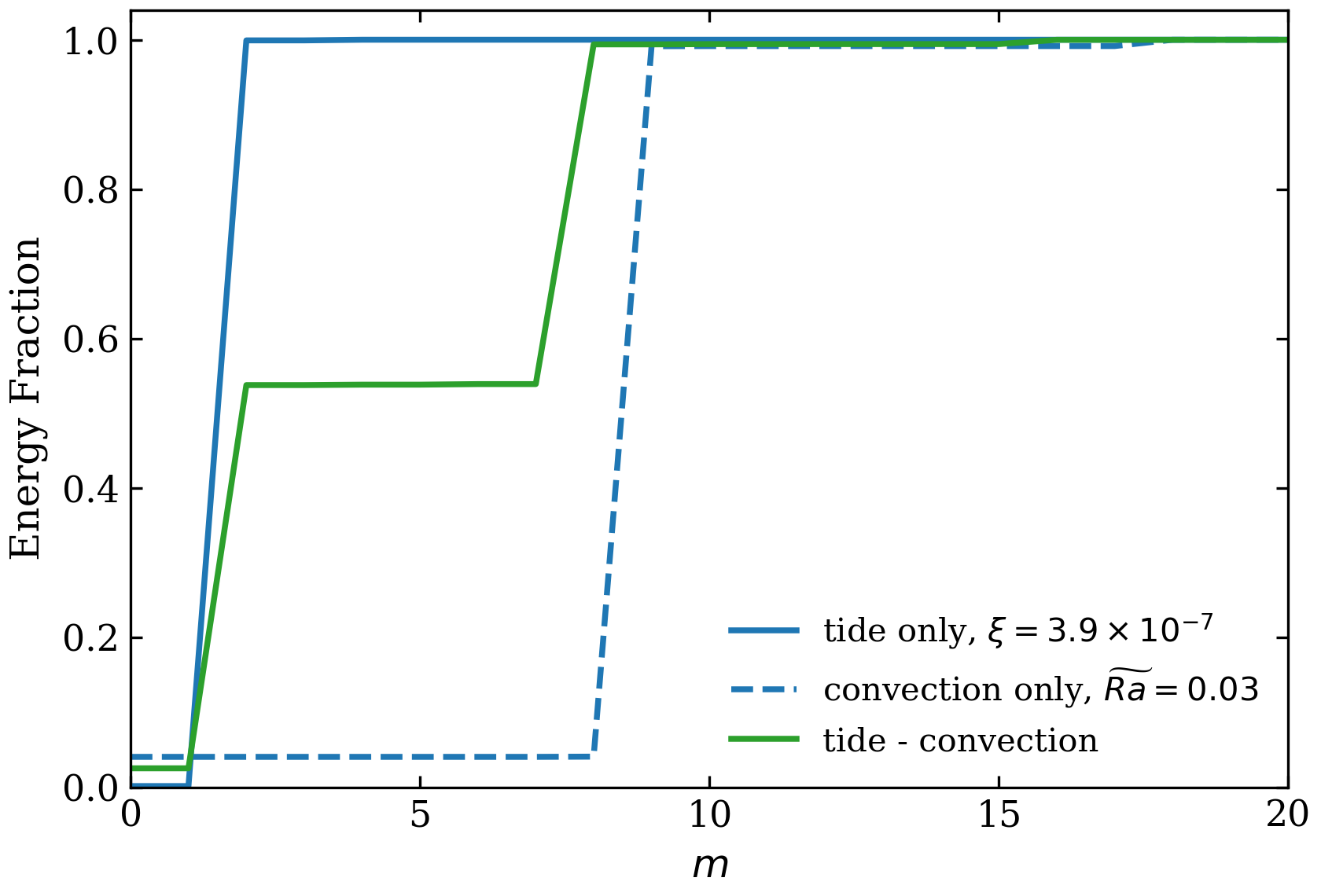}
\includegraphics[width=0.35\textwidth]{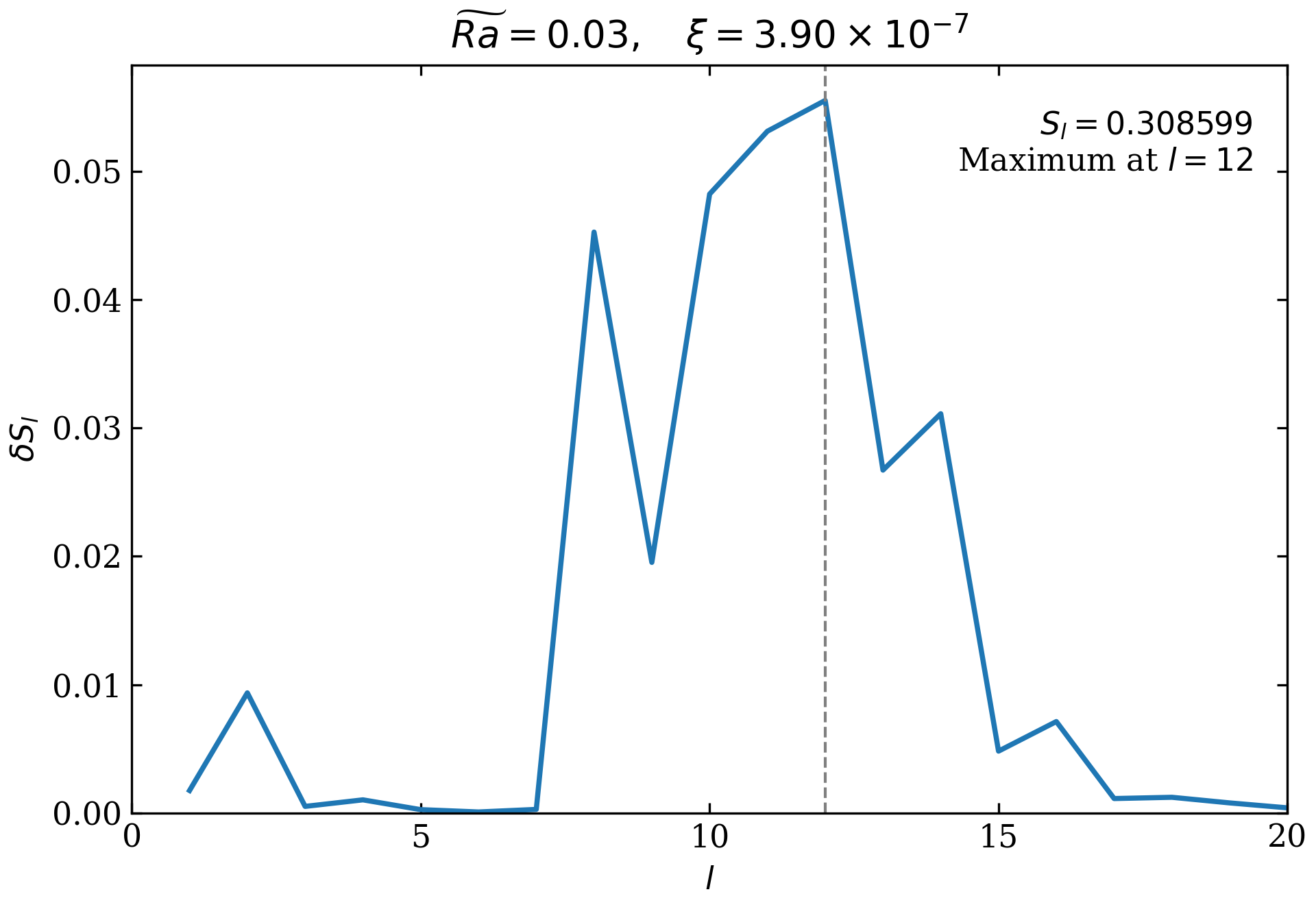}
\includegraphics[width=0.35\textwidth]{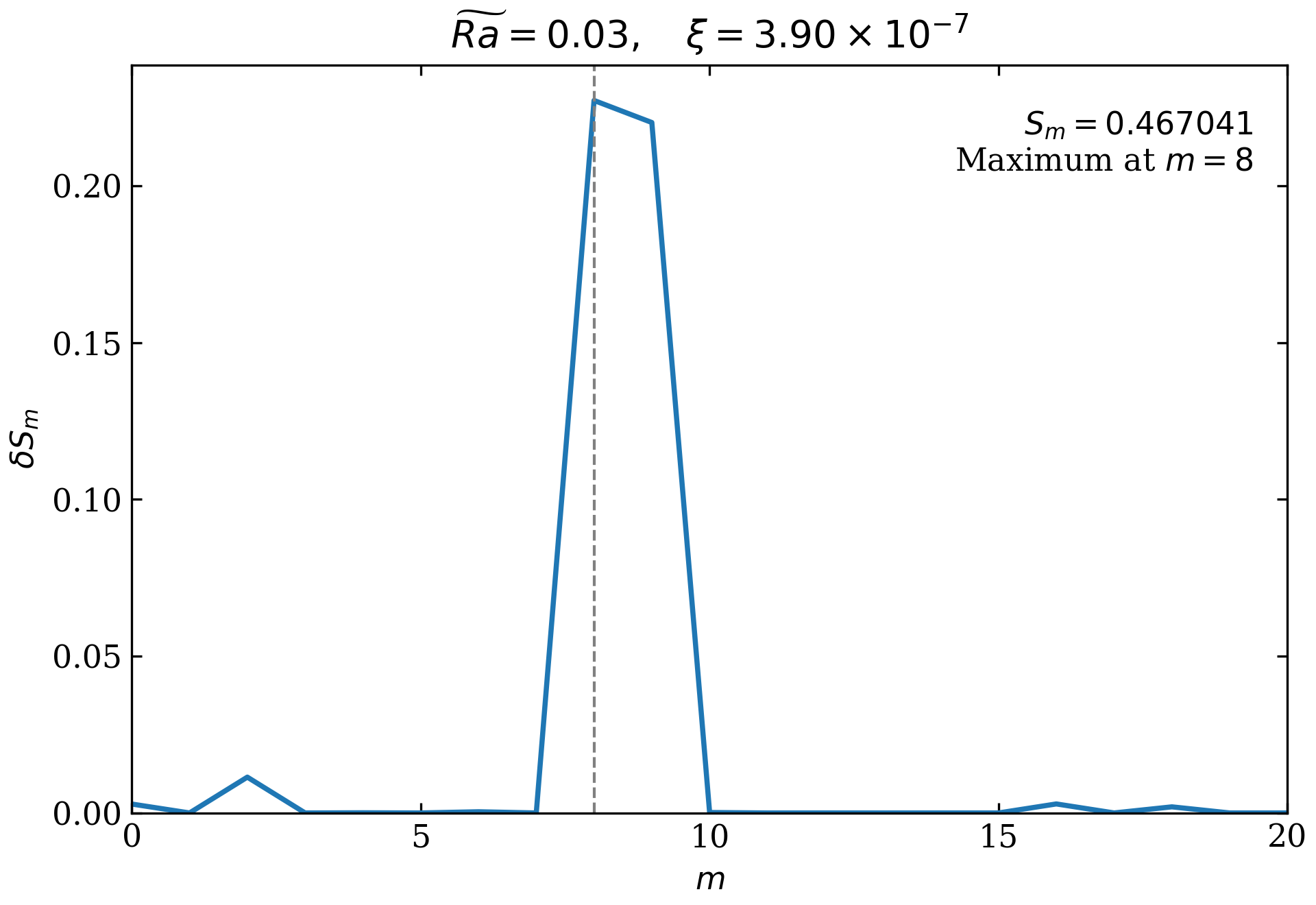}
\caption{Top panels: cumulative kinetic energy fraction as function of $l$ (left) and $m$ (right). Bottom panels: relative error of energy spectra between tide-convection flow and superposition of two pure flows, $l$ spectrum (left) and $m$ spectrum (right).}
\label{fig:fraction-difference}
\end{figure*}

To quantify the characteristic spatial scales, we calculate the cumulative kinetic energy fractions as function of spherical harmonic degree $l$ and order $m$, shown in the two top panels of Fig.~\ref{fig:fraction-difference}. Clearly, tide occupies the large scale but convection small scale, and tide-convection the intermediate. To give the boundary of energy fraction, we define $l_{95}$ or $m_{95}$ as the degree or order containing 95\% of the total kinetic energy. We find $(l_{95} = 6, m_{95} = 2)$ for tidal flow, $(l_{95} = 16, m_{95} = 9)$ for convective flow, and $(l_{95} = 13, m_{95} = 8)$ for tide-convection flow. This is qualitatively consistent with the global tidal models \citep{hay2019, hay2024} and the rotating-convection studies \citep{Soderlund2019, Pagnoscin2026}. Tidally driven shear may induce small scales through parametric instability due to resonance of inertial waves \citep{Rovira-Navarro2019}, but the most energy is contained in large scale. The tide-convection flow therefore retains at a large scale while its spectrum is broadened due to convection (see top left panel of Fig.~\ref{fig:fraction-difference}). We define $\lambda_h = 2\pi r/\sqrt{l(l+1)}$ to evaluate energy-cutoff scale at the mid-shell radius $r = 0.75 r_o$. Since $d = 0.5 r_o$, the corresponding 95\% energy-cutoff scales are approximately $1.45d$, $0.57d$, and $0.70d$ for tidal flow, convective flow, and tide-convection flow, respectively. The ocean depth magnitude of icy satellites is about 1000 km, so that the flow energy is contained in the scale of several hundred kilometres. Furthermore, we calculate the relative error of energy spectra between tide-convection flow and superposition of two pure flows, as shown in the two bottom panels of Fig.~\ref{fig:fraction-difference}. This method follows the ocean-wave analysis \citep{Alvarez-Esteban2016}. Clearly, the mode-mode transition occurs at intermediate scales with the maximum $l=12$ and $m=8$.

\begin{figure*}
\centering
\includegraphics[width=0.36\textwidth]{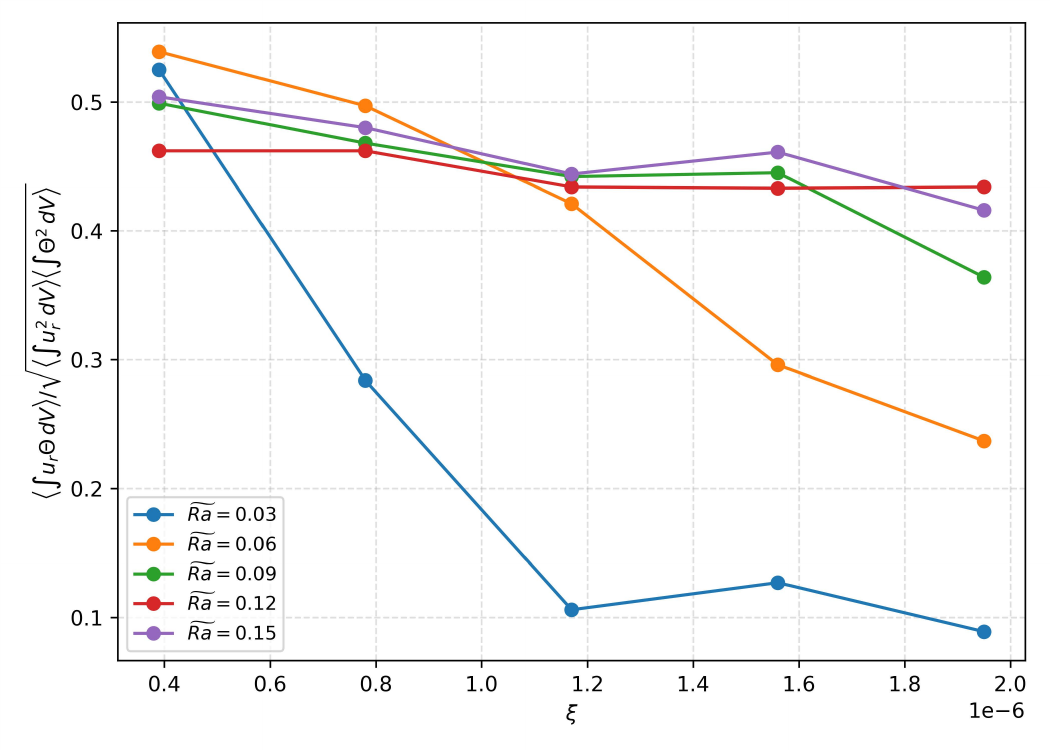}
\includegraphics[width=0.34\textwidth]{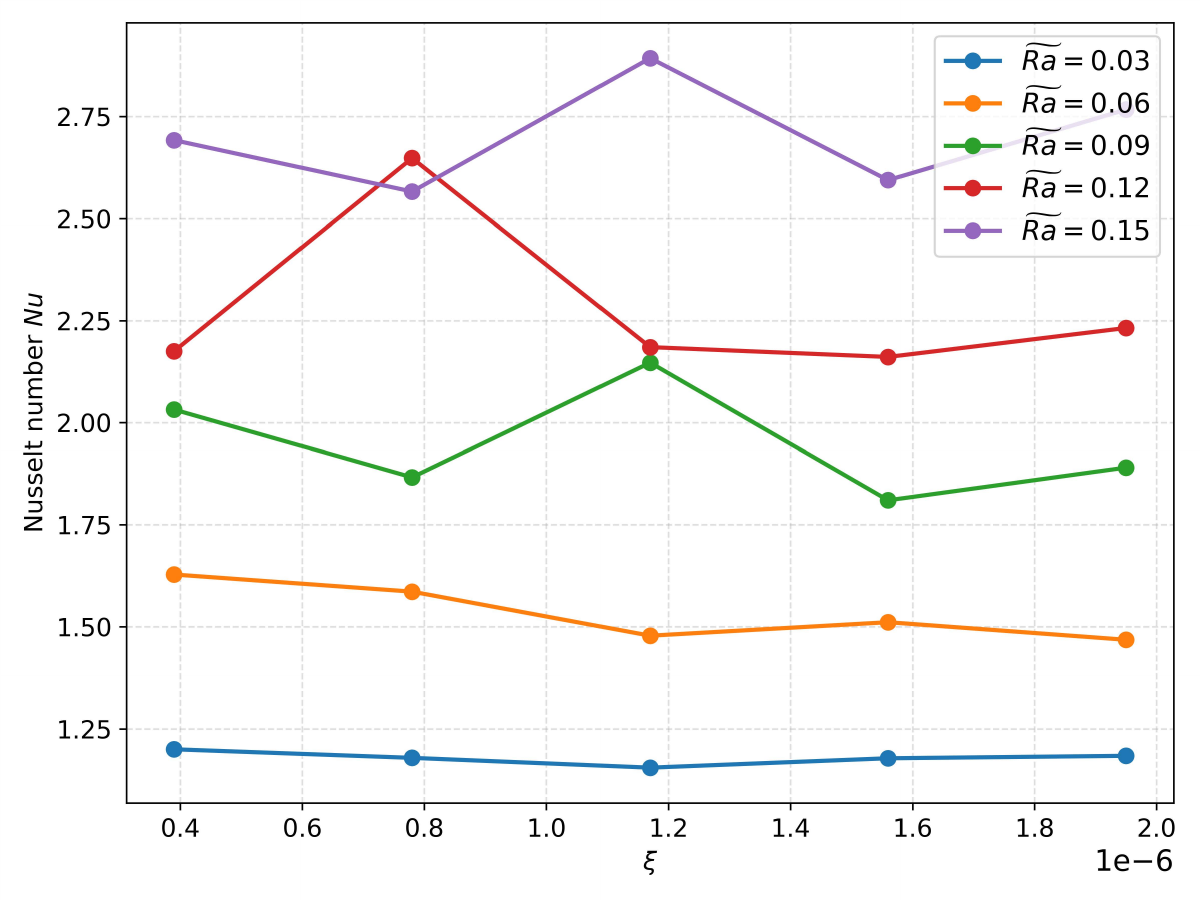}
\caption{Left panel: correlation coefficient of heat transport $\int u_r \Theta \, dV/\sqrt{\left(\int u_r^2 \, dV\right)\left(\int \Theta^2 \, dV\right)}$ versus $\xi$ at different $\widetilde{Ra}$. Right panel: Nusselt number $Nu=1-\langle\partial\Theta/\partial r\rangle$ versus $\xi$ at different $\widetilde{Ra}$.}
\label{fig:heat_transport}
\end{figure*}

To better understand how tide and convection interact to influence heat transport, we investigate the correlation coefficient of heat transport $\int u_r \Theta \, dV/\sqrt{\left(\int u_r^2 \, dV\right)\left(\int \Theta^2 \, dV\right)}$ and Nusselt number $Nu$. The correlation coefficient reflects the phase difference between radial motion and temperature perturbation, and measures the efficiency of kinetic energy converted into heat. Nusselt number is the ratio of convective heat flux to conductive heat flux and measures convective heat flux, $Nu=\langle\partial T/\partial r\rangle/(dT_b/dr)=1-\langle\partial\Theta/\partial r\rangle$ where $T=T_b+\Theta$ (see \S\ref{sec:method}) and bracket denotes the spherical surface average at any radius $r$. Because in our model there is no heat source in the interior, $Nu$ is always the same across all radii. Fig.~\ref{fig:heat_transport} shows the correlation coefficient and Nusselt number versus tide with different buoyancy force. Left panel shows that higher $\xi$ at fixed $\widetilde{Ra}$ leads to lower correlation coefficient. This indicates that tide suppresses the efficiency of heat transport. It might be because the large-scale tidal shear distorts buoyancy plumes to induce a phase shift between $u_r$ and $\Theta$ such that tidal flow is not optimally aligned with thermal structure. With stronger buoyancy force, the tidal suppression on heat transport becomes less important. Right panel shows that stronger buoyancy force leads to higher $Nu$ but the tidal effect on $Nu$ is not monotonic.

\begin{figure*}
 \centering
 \includegraphics[width=0.35\textwidth]{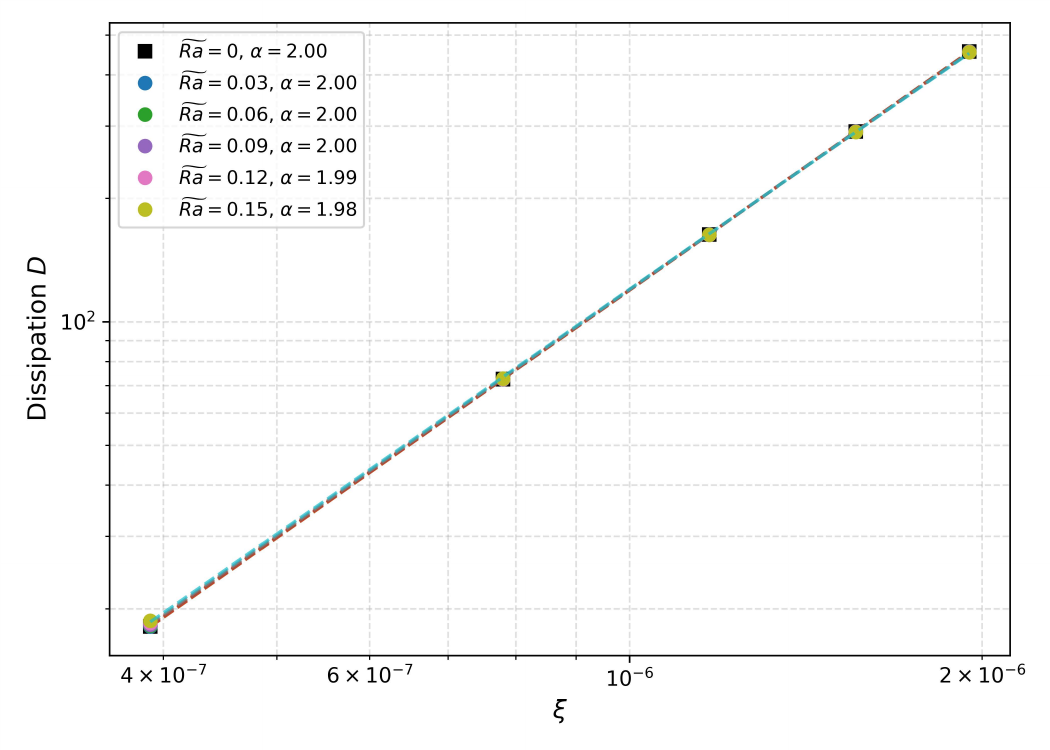}
 \includegraphics[width=0.35\textwidth]{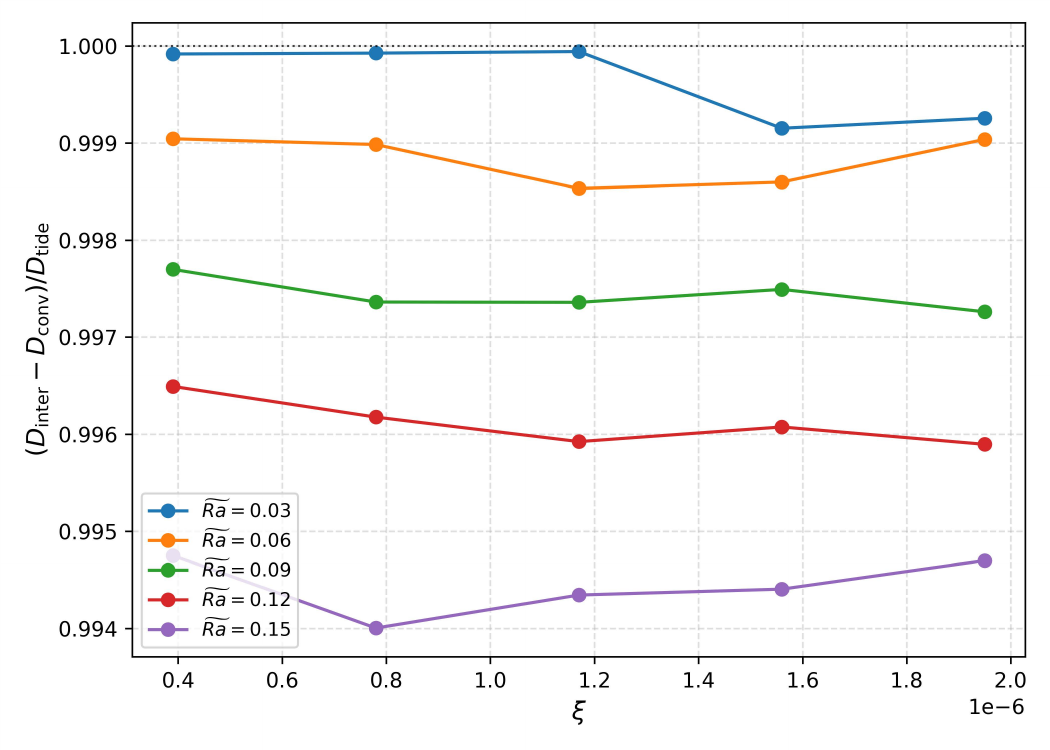}
 \caption{Quantities about dissipation versus $\xi$ at different $\widetilde{Ra}$. Left panel: time-averaged dissipation. Right panel: coefficient  $(D_{\rm inter}-D_{\rm conv})/D_{\rm tide}$.}
 \label{fig:dissipation}
\end{figure*}

To end this section, we investigate the dissipation of kinetic energy. The dissipation in dimensionless equations is calculated as $\int|\nabla\times \mathbf{u}|^2dV$. Fig.~\ref{fig:dissipation} shows the quantities about dissipation versus tide with different buoyancy force. Left panel shows the scaling law of time-averaged dissipation, $D\propto \xi^2$ with a very weak dependence on $\widetilde{Ra}$. This suggests that tide controls dissipation and buoyancy force has little impact on dissipation. Right panel shows $(D_{\rm inter}-D_{\rm conv})/D_{\rm tide}$, i.e., the additional dissipation when tide is added to convection normalised by the tide-only dissipation. This coefficient exhibits how tide influences dissipation. It is always less than 1, and stronger buoyancy force leads to lower coefficient, which implies that buoyancy force suppresses the tidal influence on dissipation.

\begin{comment}
To end this section we investigate the equivalence between tidal dissipation and convective heat. The dimensional dissipation is $\rho r_o^5\Omega^3 Ek D$ where $D$ is dimensionless dissipation. The dimensional convective heat $4\pi\rho c_p\Delta Tr_o^3\Omega(Ek/Pr)Nu$ is proportional to Nusselt number, where $c_p$ is specific heat capacity. Therefore, the ratio of dimensional tidal dissipation to dimensional convective heat is $r_o^2\Omega^2/(4\pi c_p\Delta T)Pr(D/Nu)$. Right panel of Fig.~\ref{fig:dissipation} shows the ratio $D/Nu$. It is not surprising that stronger tide leads to larger ratio (because of larger $D$) but stronger buoyancy force leads to smaller ratio (because of larger $Nu$). We can find this trend but cannot answer the equivalence because the realistic parameters about convection, such as $\Delta T$ and $Nu$, are unknown.
\end{comment}

\section{Discussion}\label{sec:discussion}

In this short paper we try to measure the interaction of eccentricity tide and convection in the subsurface ocean of icy satellites. We find the scaling laws of energy and dissipation, and how this interaction influences the energy, spectra, heat transport, dissipation, etc. We also find that tide controls large scale of fluid motion but convection is important for small scale, and the large-scale tidal shear suppresses the efficiency of heat transport. These results are of help to understanding the fluid motion and heat transport in icy satellites. 

We now make an order-of-magnitude estimation for synchronous icy satellites. The potential of eccentricity tide is roughly $\Psi \sim en^2R^2$ where $e$ is orbital eccentricity, $n$ mean motion and $R$ satellite radius. The tidal flow is then roughly $u_r \sim n\Psi/g$ where $g$ is surface gravity acceleration. The dimensionless tidal flow is roughly $\xi \sim u_r/(nR) \sim e(M/m)(R/a)^3$ where $M$ is planet mass, $m$ satellite mass and $a$ semi-major axis. On the other hand, the dimensionless $\widetilde{Ra} \sim \alpha\beta g/n^2 \sim \alpha(h/\lambda)(m/M)R$ where $\alpha$ is thermal expansivity, $\beta$ magnitude of temperature gradient, $h$ heat flux and $\lambda$ thermal conductivity. Table~\ref{tab:estimation} shows the estimation, where the estimation of heat flux is from \citep{kang2023modulation, Lemasquerier2023, Schubert2004, vance2018geophysical, spohn2003oceans}. Our $\xi$ is in the realistic regime. Traditional Rayleigh number and our rotationally modified Rayleigh number are related by $Ra=\widetilde{Ra}E^{-2}Pr$. Our Ekman number $2\times10^{-4}$ is much higher than the realistic value $\sim 10^{-13}$ (radius $R\sim 10^3\;{\rm km}$, rotation rate $\Omega\sim 10^{-5}\;{\rm s}^{-1}$ and viscosity $10^{-6}\;{\rm m^2/s}$). Our Prandtl number $Pr=1$ is comparable to the realistic value $\sim 10$. Then our traditional Rayleigh number $Ra\sim 10^6-10^7$ is much lower than the realistic value $Ra\sim 10^{19}-10^{23}$. Tidal suppression on heat transport and tidal contribution to dissipation need to be explored in the highly supercritical regime.

\begin{table*}
\centering
\caption{Estimation of $\xi$ and $\widetilde{Ra}$ for synchronous icy satellites. Thermal expansivity $\alpha=10^{-4}\ {\rm K^{-1}}$ \citep{kang2023modulation} and thermal conductivity $\lambda=0.6\ {\rm W\,m^{-1}}\,K^{-1}$.}
\label{tab:estimation}
\begin{tabular}{lccccc}
\hline
  & Europa & Ganymede & Callisto & Enceladus & Titan \\
\hline
$M/m$ & $3.96\times10^{4}$ & $1.28\times10^{4}$ & $1.76\times10^{4}$ & $5.26\times10^{6}$ & $4.22\times10^{3}$ \\
$a$ (km) & $6.71\times10^{5}$ & $1.07\times10^{6}$ & $1.88\times10^{6}$ & $2.38\times10^{5}$ & $1.22\times10^{6}$ \\
$R$ (km) & $1.56\times10^{3}$ & $2.63\times10^{3}$ & $2.41\times10^{3}$ & $2.52\times10^{2}$ & $2.57\times10^{3}$ \\
$e$ & 0.009 & 0.001 & 0.007 & 0.005 & 0.029 \\
$Ek$ & $4.11\times10^{-14}$ & $1.45\times10^{-14}$ & $1.72\times10^{-14}$ & $1.57\times10^{-12}$ & $1.51\times10^{-14}$ \\
$h$ (mW/m$^2$) & 6-46 & 15-107 & 3-7 & 16-83 & 14-20 \\
$\xi$ & $4.48\times10^{-6}$ & $1.90\times10^{-7}$ & $2.60\times10^{-7}$ & $3.11\times10^{-5}$ & $1.15\times10^{-6}$ \\
$\widetilde{Ra}$ & $(3.96\text{--}30.2)\times10^{-5}$ & $(5.15\text{--}36.8)\times10^{-4}$ & $(6.82\text{--}15.9)\times10^{-5}$ & $(1.28\text{--}6.62)\times10^{-7}$ & $(1.42\text{--}2.03)\times10^{-3}$ \\
\hline
\end{tabular}
\end{table*}

More work needs to be explored in the future. Firstly, analytical multiscale calculation about triad resonance in Cartesian geometry, similar to \citet{kerswell2002jfm}, is needed to understand the detail about the interaction of tide and convection. Secondly, more numerical simulations need to be performed in a more highly supercritical regime at lower Ekman number and higher Rayleigh number to find better scaling laws of energy and dissipation. Thirdly, \citet{goldreich-lithwick2025} studied the limit cycle of Enceladus, i.e., coupling of orbit and internal structure leads to the cycles of ice freezing and melting. Our work may put some new ingredient. During the limit cycle, the interaction of eccentricity tide and convection should be taken into account. But how to model this additional effect in limit cycle needs to be further explored.

\bibliographystyle{aa}
\bibliography{paper}

@article{nimmo2016ocean,
  title={Ocean worlds in the outer solar system},
  author={Nimmo, F and Pappalardo, Robert T},
  journal={Journal of Geophysical Research: Planets},
  volume={121},
  number={8},
  pages={1378--1399},
  year={2016},
  publisher={Wiley Online Library}
}

@INCOLLECTION{nimmo2009geodynamics,
       author = {{Nimmo}, F. and {Manga}, M.},
        title = "{Geodynamics of Europa's Icy Shell}",
    booktitle = {Europa},
         year = 2009,
       editor = {{Pappalardo}, Robert T. and {McKinnon}, William B. and {Khurana}, Krishan K.},
        pages = {381},
    publisher = {University of Arizona Press},
       adsurl = {https://ui.adsabs.harvard.edu/abs/2009euro.book..381N}
}

@article{chen2014tidal,
  title={Tidal heating in icy satellite oceans},
  author={Chen, EMA and Nimmo, Francis and Glatzmaier, Gary A},
  journal={Icarus},
  volume={229},
  pages={11--30},
  year={2014},
  publisher={Elsevier}
}

@article{kang2022different,
  title={Different ice-shell geometries on Europa and Enceladus due to their different sizes: impacts of ocean heat transport},
  author={Kang, Wanying},
  journal={The Astrophysical Journal},
  volume={934},
  number={2},
  pages={116},
  year={2022},
  publisher={The American Astronomical Society}
}

@article{jansen2023energetic,
  title={Energetic constraints on ocean circulations of icy ocean worlds},
  author={Jansen, Malte F and Kang, Wanying and Kite, Edwin S and Zeng, Yaoxuan},
  journal={The Planetary Science Journal},
  volume={4},
  number={6},
  pages={117},
  year={2023},
  publisher={The American Astronomical Society}
}

@article{kang2023modulation,
  title={The modulation effect of ice thickness variations on convection in icy ocean worlds},
  author={Kang, Wanying},
  journal={Monthly Notices of the Royal Astronomical Society},
  volume={525},
  number={4},
  pages={5251--5261},
  year={2023},
  publisher={Oxford University Press}
}

@article{ogilvie2014tidal,
  title={Tidal dissipation in stars and giant planets},
  author={Ogilvie, Gordon I},
  journal={Annual Review of Astronomy and Astrophysics},
  volume={52},
  number={1},
  pages={171--210},
  year={2014},
  publisher={Annual Reviews}
}

@article{barker2021interaction,
  title={On the interaction between fast tides and convection},
  author={Barker, Adrian J and Astoul, Aur{\'e}lie AV},
  journal={Monthly Notices of the Royal Astronomical Society: Letters},
  volume={506},
  number={1},
  pages={L69--L73},
  year={2021},
  publisher={Oxford University Press}
}

@article{goodman1997fast,
  title={Fast tides in slow stars: the efficiency of eddy viscosity},
  author={Goodman, Jeremy and Oh, Siang Peng},
  journal={The Astrophysical Journal},
  volume={486},
  number={1},
  pages={403--412},
  year={1997}
}

@article{penev2009dissipation,
  title={Dissipation efficiency in turbulent convective zones in low-mass stars},
  author={Penev, Kaloyan and Sasselov, Dimitar and Robinson, Frank and Demarque, Pierre},
  journal={The Astrophysical Journal},
  volume={704},
  number={2},
  pages={930--936},
  year={2009},
  publisher={The American Astronomical Society}
}

@article{goldreich1977turbulent,
  title={Turbulent viscosity and Jupiter's tidal Q},
  author={Goldreich, Peter and Nicholson, Philip D},
  journal={Icarus},
  volume={30},
  number={2},
  pages={301--304},
  year={1977},
  publisher={Elsevier}
}

@article{zahn1977tidal,
  title={Tidal friction in close binary stars},
  author={Zahn, J-P},
  journal={Astronomy and Astrophysics, vol. 57, no. 3, May 1977, p. 383-394.},
  volume={57},
  pages={383--394},
  year={1977}
}

@article{hollerbach2000spectral,
  title={A spectral solution of the magneto-convection equations in spherical geometry},
  author={Hollerbach, Rainer},
  journal={International journal for numerical methods in fluids},
  volume={32},
  number={7},
  pages={773--797},
  year={2000},
  publisher={Wiley Online Library}
}

@article{wei2013stratified,
  title={Stratified precessional flow in spherical geometry},
  author={Wei, Xing and Tilgner, Andreas},
  journal={Journal of Fluid Mechanics},
  volume={718},
  pages={R2},
  year={2013},
  publisher={Cambridge University Press}
}

@article{de2025tidal,
  title={Tidal dissipation and spin--orbit alignment due to the precessional instability in convection zones in rotating giant planets and stars},
  author={de Vries, Nils B and Barker, Adrian J and Hollerbach, Rainer},
  journal={Monthly Notices of the Royal Astronomical Society},
  volume={544},
  number={4},
  pages={3324--3349},
  year={2025},
  publisher={Oxford University Press}
}

@BOOK{chandrasekhar1961,
       author = {{Chandrasekhar}, Subrahmanyan},
        title = "{Hydrodynamic and hydromagnetic stability}",
         year = 1961,
    publisher = {Oxford Clarendon}
}

@article{lithwick2025enceladus,
  title={Enceladus’s Tidal Heating: A Simple Approach},
  author={Lithwick, Yoram},
  journal={The Planetary Science Journal},
  volume={6},
  number={10},
  pages={227},
  year={2025},
  publisher={The American Astronomical Society}
}

@article{lebars2015flows,
  title={Flows driven by libration, precession, and tides},
  author={Le Bars, Michael and C{\'e}bron, David and Le Gal, Patrice},
  journal={Annual Review of Fluid Mechanics},
  volume={47},
  number={1},
  pages={163--193},
  year={2015},
  publisher={Annual Reviews}
}

@ARTICLE{wei2018,
       author = {{Wei}, Xing},
        title = "{The Magnetic Effect on Dynamical Tide in Rapidly Rotating Astronomical Objects}",
      journal = {Astrophys. J.},
         year = 2018,
       volume = {854},
        pages = {34}
}

@article{kerswell1993,
  title={The instability of precessing flow},
  author={Kerswell, RR},
  journal={Geophysical \& Astrophysical Fluid Dynamics},
  volume={72},
  number={1-4},
  pages={107--144},
  year={1993},
  publisher={Taylor \& Francis}
}

@ARTICLE{barker2022,
       author = {{Astoul}, A. and {Barker}, A.~J.},
        title = "{The effects of non-linearities on tidal flows in the convective envelopes of rotating stars and planets in exoplanetary systems}",
      journal = {\mnras},
         year = 2022,
       volume = {516},
       number = {2},
        pages = {2913-2935}
}

@ARTICLE{kerswell2002jfm,
       author = {{Mason}, R.~M. and {Kerswell}, R.~R.},
        title = "{Chaotic dynamics in a strained rotating flow: a precessing plane fluid layer}",
      journal = {Journal of Fluid Mechanics},
         year = 2002,
       volume = {471},
        pages = {71-106}
}

@ARTICLE{Lemasquerier2023,
       author = {{Lemasquerier}, D.~G. and {Bierson}, C.~J. and {Soderlund}, K.~M.},
        title = "{Europa's Ocean Translates Interior Tidal Heating Patterns to the Ice-Ocean Boundary}",
      journal = {AGU Advances},
         year = 2023,
        month = dec,
       volume = {4},
       number = {6},
          eid = {e2023AV000994},
        pages = {e2023AV000994}
}

@INCOLLECTION{Schubert2004,
       author = {{Schubert}, G. and {Anderson}, J.~D. and {Spohn}, T. and {McKinnon}, W.~B.},
        title = "{Interior composition, structure and dynamics of the Galilean satellites}",
    booktitle = {Jupiter. The Planet, Satellites and Magnetosphere},
         year = 2004,
       editor = {{Bagenal}, Fran and {Dowling}, Timothy E. and {McKinnon}, William B.},
       volume = {1},
        pages = {281-306},
    publisher = {Cambridge University Press}
}

@article{vance2018geophysical,
  title={Geophysical investigations of habitability in ice-covered ocean worlds},
  author={Vance, Steven D and Panning, Mark P and St{\"a}hler, Simon and Cammarano, Fabio and Bills, Bruce G and Tobie, Gabriel and Kamata, Shunichi and Kedar, Sharon and Sotin, Christophe and Pike, William T and others},
  journal={Journal of Geophysical Research: Planets},
  volume={123},
  number={1},
  pages={180--205},
  year={2018},
  publisher={Wiley Online Library}
}

@article{spohn2003oceans,
  title={Oceans in the icy Galilean satellites of Jupiter?},
  author={Spohn, Tilman and Schubert, Gerald},
  journal={Icarus},
  volume={161},
  number={2},
  pages={456--467},
  year={2003},
  publisher={Elsevier}
}

@article{lin2026,
  title={The instability of precession-driven flows in a stably stratified spherical shell},
  author={Cheng, Rui and Fan, Wei and Lin, Yufeng},
  journal={Journal of Fluid Mechanics},
  year={2026},
  volume={1030},
  number={2},
  pages={A62}
}

@ARTICLE{Vidal2020apjl,
       author = {{Vidal}, J{\'e}r{\'e}mie and {Barker}, Adrian J.},
        title = "{Turbulent Viscosity Acting on the Equilibrium Tidal Flow in Convective Stars}",
      journal = {\apjl},
         year = 2020,
        month = jan,
       volume = {888},
       number = {2},
          eid = {L31},
        pages = {L31},
          doi = {10.3847/2041-8213/ab6219},
archivePrefix = {arXiv},
       eprint = {1912.07910},
 primaryClass = {astro-ph.SR},
       adsurl = {https://ui.adsabs.harvard.edu/abs/2020ApJ...888L..31V}
}

@ARTICLE{Vidal2020MNRAS,
       author = {{Vidal}, J{\'e}r{\'e}mie and {Barker}, Adrian J.},
        title = "{Efficiency of tidal dissipation in slowly rotating fully convective stars or planets}",
      journal = {\mnras},
         year = 2020,
        month = oct,
       volume = {497},
       number = {4},
        pages = {4472-4485},
          doi = {10.1093/mnras/staa2239},
archivePrefix = {arXiv},
       eprint = {2007.13392},
 primaryClass = {astro-ph.SR},
       adsurl = {https://ui.adsabs.harvard.edu/abs/2020MNRAS.497.4472V}
}

@ARTICLE{Vidal2023JFM,
       author = {{Vidal}, J{\'e}r{\'e}mie and {C{\'e}bron}, David},
        title = "{Precession-driven flows in stress-free ellipsoids}",
      journal = {Journal of Fluid Mechanics},
         year = 2023,
        month = jan,
       volume = {954},
          eid = {A5},
        pages = {A5},
          doi = {10.1017/jfm.2022.976},
archivePrefix = {arXiv},
       eprint = {2301.03254},
 primaryClass = {physics.flu-dyn},
       adsurl = {https://ui.adsabs.harvard.edu/abs/2023JFM...954A...5V}
}

@article{terquem2026turbulent,
  title={Turbulent damping of fast tidal oscillations by three-dimensional Rayleigh-B{\'e}nard convection with a radiating free surface},
  author={Terquem, Caroline and Martinez, Enrico and Boone, Alexander},
  journal={Monthly Notices of the Royal Astronomical Society},
  pages={stag909},
  year={2026},
  publisher={Oxford University Press}
}

@ARTICLE{goldreich-lithwick2025,
       author = {{Goldreich}, Peter and {Lithwick}, Yoram and {Luan}, Jing},
        title = "{Enceladus's Limit Cycle}",
      journal = {\apj},
         year = 2025,
        month = oct,
       volume = {992},
       number = {1},
          eid = {28},
        pages = {28},
          doi = {10.3847/1538-4357/adf8e7},
archivePrefix = {arXiv},
       eprint = {2503.01967},
 primaryClass = {astro-ph.EP},
       adsurl = {https://ui.adsabs.harvard.edu/abs/2025ApJ...992...28G}
}

@ARTICLE{hartmann2024,
       author = {{Hartmann}, Robert and {Stevens}, Richard J.~A.~M. and {Lohse}, Detlef and {Verzicco}, Roberto},
        title = "{Toward Understanding Polar Heat Transport Enhancement in Subglacial Oceans on Icy Moons}",
      journal = {\grl},
         year = 2024,
        month = feb,
       volume = {51},
       number = {3},
          eid = {e2023GL105401},
        pages = {e2023GL105401},
          doi = {10.1029/2023GL105401},
       adsurl = {https://ui.adsabs.harvard.edu/abs/2024GeoRL..5105401H}
}

@ARTICLE{lin2024,
       author = {{Fan}, Wei and {Wang}, Qi and {Lin}, Yufeng},
        title = "{Scaling behaviour of rotating convection in a spherical shell with different Prandtl numbers}",
      journal = {Journal of Fluid Mechanics},
         year = 2024,
        month = nov,
       volume = {998},
          eid = {A20},
        pages = {A20},
          doi = {10.1017/jfm.2024.893},
archivePrefix = {arXiv},
       eprint = {2508.09416},
 primaryClass = {physics.flu-dyn},
       adsurl = {https://ui.adsabs.harvard.edu/abs/2024JFM...998A..20F}
}

@ARTICLE{hay2019,
       author = {{Hay}, Hamish C.~F.~C. and {Matsuyama}, Isamu},
        title = "{Nonlinear tidal dissipation in the subsurface oceans of Enceladus and other icy satellites}",
      journal = {\icarus},
         year = 2019,
        month = feb,
       volume = {319},
        pages = {68-85},
          doi = {10.1016/j.icarus.2018.09.019},
       adsurl = {https://ui.adsabs.harvard.edu/abs/2019Icar..319...68H}
}

@ARTICLE{hay2024,
       author = {{Hay}, Hamish C.~F.~C. and {Hewitt}, Ian and {Katz}, Richard F.},
        title = "{Tidal Forcing in Icy-Satellite Oceans Drives Mean Circulation and Ice-Shell Torques}",
      journal = {Journal of Geophysical Research (Planets)},
         year = 2024,
        month = jun,
       volume = {129},
       number = {6},
          eid = {e2024JE008408},
        pages = {e2024JE008408},
          doi = {10.1029/2024JE008408},
       adsurl = {https://ui.adsabs.harvard.edu/abs/2024JGRE..12908408H}
}

@ARTICLE{Pagnoscin2026,
       author = {{Pagnoscin}, Silvia and {von Hardenberg}, Jost and {Brucato}, John Robert and {Provenzale}, Antonello},
        title = "{Convection in the subsurface ocean of icy moons and response of the upper ice layer}",
      journal = {\icarus},
         year = 2026,
        month = mar,
       volume = {446},
          eid = {116875},
        pages = {116875},
          doi = {10.1016/j.icarus.2025.116875},
       adsurl = {https://ui.adsabs.harvard.edu/abs/2026Icar..44616875P}
}

@ARTICLE{Rovira-Navarro2019,
       author = {{Rovira-Navarro}, Marc and {Rieutord}, Michel and {Gerkema}, Theo and {Maas}, Leo R.~M. and {van der Wal}, Wouter and {Vermeersen}, Bert},
        title = "{Do tidally-generated inertial waves heat the subsurface oceans of Europa and Enceladus?}",
      journal = {\icarus},
         year = 2019,
        month = mar,
       volume = {321},
        pages = {126-140},
          doi = {10.1016/j.icarus.2018.11.010},
       adsurl = {https://ui.adsabs.harvard.edu/abs/2019Icar..321..126R}
}

@ARTICLE{Soderlund2019,
       author = {{Soderlund}, Krista M.},
        title = "{Ocean Dynamics of Outer Solar System Satellites}",
      journal = {\grl},
         year = 2019,
        month = aug,
       volume = {46},
       number = {15},
        pages = {8700-8710},
          doi = {10.1029/2018GL081880},
archivePrefix = {arXiv},
       eprint = {1901.04093},
 primaryClass = {physics.geo-ph},
       adsurl = {https://ui.adsabs.harvard.edu/abs/2019GeoRL..46.8700S}
}

@ARTICLE{Alvarez-Esteban2016,
       author = {{Alvarez-Esteban}, Pedro C. and {Eu{\'a}n}, Carolina and {Ortega}, Joaqu{\'\i}n},
        title = "{Time series clustering using the total variation distance with applications in oceanography}",
      journal = {Environmetrics},
         year = 2016,
        month = sep,
       volume = {27},
       number = {6},
        pages = {355-369},
          doi = {10.1002/env.2398},
archivePrefix = {arXiv},
       eprint = {1501.04050},
 primaryClass = {stat.ME},
       adsurl = {https://ui.adsabs.harvard.edu/abs/2016Envir..27..355A}
}

\end{document}